\documentclass[12pt]{article}
\usepackage{amsmath,latexsym,amssymb,graphics,theorem}
\usepackage[pdftex]{graphicx}
\usepackage{amsmath}

\usepackage{epstopdf}

\newtheorem{teo}{Theorem}[section]

{\theoremstyle{plain}\theoremheaderfont{\bfseries\itshape}
\newtheorem{prop}[teo]{Proposition}}

\newtheorem{lemma}[teo]{Lemma}

{\theoremstyle{plain}\theorembodyfont{\rmfamily}
\theoremheaderfont{\itshape} \newtheorem{remark}[teo]{Remark}}

\numberwithin{equation}{section}

\long\def\symbolfootnote[#1]#2{\begingroup%
\def\thefootnote{\fnsymbol{footnote}}\footnote[#1]{#2}\endgroup}

\newcommand{\R}{\mathbb{R}}

\newcommand{\scri}{{\cal J}}
\newcommand{\Par}{\par}

\newcommand\calS{{\mathcal{S}}}

\newcommand\g{\gamma}

\newcommand\beq{\begin{equation}}
\newcommand\eeq{\end{equation}}
\newcommand\ben{\begin{enumerate}}
\newcommand\een{\end{enumerate}}
\newcommand\bit{\begin{itemize}}
\newcommand\eit{\end{itemize}}

\newcounter{mnotecount}[section]

\title{Uniqueness of de Sitter Space}

\author{
Gregory J. Galloway
\\
Department of Mathematics \\ University of Miami, Coral Gables, FL
33124, U.S.A. \\ \\
Didier A. Solis \\
Facultad de Matem\'aticas \\ Universidad Aut\'onoma de Yucat\'an,
M\'erida, M\'exico }

\begin{document}
\date{}
\maketitle

\begin{abstract}
All inextendible null geodesics in four dimensional de Sitter space
$dS^4$ are complete and globally achronal.  This achronality is
related to the fact that all observer horizons in $dS^4$ are
eternal, i.e. extend from future infinity $\scri^+$ all the way back
to past infinity $\scri^-$.  We show that the property of having a
null line (inextendible achronal null geodesic) that extends from
$\scri^-$ to $\scri^+$ characterizes $dS^4$ among all globally
hyperbolic and asymptotically de Sitter spacetimes satisfying the
vacuum Einstein equations with positive cosmological constant. This
result is then further extended to allow for a class of matter
models  that includes perfect fluids.
\end{abstract}

\section{Introduction}
Asymptotically de Sitter spacetimes can be roughly thought of as
solutions to the Einstein equations with positive cosmological
constant having a spacelike boundary at infinity $\scri$. These
spacetimes naturally arise in a number of contexts, such as in the
study of inflationary cosmological models.  An asymptotically de
Sitter spacetime is  said to be asymptotically simple provided
every null geodesic extends all the way from past infinity
$\scri^-$ to future infinity $\scri^+$. Such spacetimes are, of
course, modelled on de Sitter space $dS^n$
itself, which conformally embeds into
the Einstein cylinder, acquiring there a past conformal  infinity
$\scri^-$ and future conformal infinity $\scri^+$, each spacelike
and diffeomorphic to the $(n-1)$-sphere. An additional causal
feature of de Sitter space is that every inextendible null
geodesic in it is globally achronal, i.e., never enters into its
own chronological future or past. Such null geodesics are referred
to as {\it null lines}.

As it turns out, the occurrence of null lines is a very particular
feature of de Sitter space. In $\cite{gal02}$ it is proved that
this property characterizes $dS^4$ among all four dimensional
asymptotically simple and de Sitter spacetimes:

\begin{teo}\label{rigds}
Let $(\tilde{M},\tilde{g})$ be an asymptotically simple and de
Sitter spacetime of dimension $n=4$ that satisfies the vacuum
Einstein equations with positive cosmological constant. If
$\tilde{M}$ contains a null line, then $\tilde{M}$ is isometric to
de Sitter space $dS^4$.
\end{teo}

As discussed in \cite{gal02,gal03}, this theorem can be
interpreted in terms of the initial value problem in the following
way: Friedrich's work   $\cite{F2}$ on the nonlinear stability of
de Sitter space shows that the set of asymptotically simple
solutions to the Einstein equations with positive cosmological
constant is open in the set of all maximal globally hyperbolic
solutions with compact spatial sections. As a consequence, by
slightly perturbing the initial data on a fixed Cauchy surface of
$dS^4$ we get in general an asymptotically simple solution of the
Einstein equations different from $dS^4$. Thus by virtue of
theorem $\ref{rigds}$, such a spacetime \textit{has no null
lines}. In other words, a small generic perturbation of the initial data
destroys \textit{all} null lines. This suggests that the so-called
generic condition of singularity theory \cite{HE} is in fact
generic with respect to perturbations of the initial data.

Alternatively, we could say that no other asymptotically simple
solution of the Einstein equations besides $dS^4$ develops
\textit{eternal observer horizons}. By definition, an observer
horizon ${\cal A}$ is the past achronal boundary $\partial
I^-(\gamma )$ of a future inextendible timelike curve $\g$, thus
${\cal A}$ is ruled by future inextendible achronal null
geodesics. As follows from previous comments, in the case of de
Sitter space, observer horizons are eternal, that is, all null
generators of ${\cal A}$ extend from $\scri^+$ all the way back to
$\scri^-$.

Since the observer horizon is the boundary of the region of
spacetime that can be observed by $\gamma$, the question arises as
to whether at one point $\gamma$ would be able to observe the
whole of space. More precisely, we want to know if there exists
$q\in \tilde{M}$ such that  $I^-(q)$ would contain a Cauchy
surface of spacetime. Gao and Wald were able to answer this
question affirmatively for globally hyperbolic spacetimes with
compact Cauchy surfaces, assuming null geodesic completeness, the
null energy condition and the null generic condition $\cite{GW}$.
Thus, as expressed by Bousso \cite{bou}, asymptotically de Sitter spacetimes
satisfying the conditions of the Gao and Wald result, have Penrose diagrams
that are  ``tall" compared to de Sitter space.
\footnote{Refer also to \cite{bou} for a discussion on the
relationship between the existence of eternal observer horizons
and entropy bounds on asymptotically de Sitter spacetimes.}

Though no set of the form $I^-(q)$ in $dS^4$ contains a Cauchy
surface, $I^-(q)$ gets arbitrarily close to doing so as
$q\to\scri^+$. However, notice that de Sitter space is not a
counterexample to Gao and Wald's result, since $dS^4$ does not
satisfy the null generic condition. Actually, the latter remark
enables us to interpret theorem $\ref{rigds}$ as a rigid version
of the Gao and Wald result  in the asymptotically simple (and
vacuum) context: by dropping the null generic hypothesis in
$\cite{GW}$ the conclusion will only fail if
$(\tilde{M},\tilde{g})$ is isometric to $dS^4$.

The aim of the present paper is to show that two of the basic assumptions in
Theorem \ref{rigds} can be substantially weakened.  Firstly, {\it asymptotic
simplicity} is a stringent global condition that rules out from the onset the
possible presence of singularities and black holes; examples such as
Schwarzschild de Sitter spacetime never enter the discussion.  In section 3
we show that, provided there is a null line that extends from $\scri^-$ to
$\scri^+$, the assumption of asymptotic simplicity can be replaced by the
much milder assumption of global hyperbolicity,  thus allowing a priori the occurrence of
singularities and black holes.

In precise terms, we show

\begin{teo}\label{mainteo}
Let $(\tilde{M},\tilde{g})$ be a globally hyperbolic and
asymptotically de Sitter spacetime of dimension $n=4$ satisfying
the vacuum Einstein equations with positive cosmological constant.
If $\tilde{M}$ has a null line with endpoints $p\in\scri^-$,
$q\in\scri^+$ then $(\tilde{M},\tilde{g})$ is isometric to an open
subset of de Sitter space containing a Cauchy surface.
\end{teo}

In fact, as is discussed in more detail in section 3, if $(\tilde{M},\tilde{g})$
is the maximal development of initial data from one of its  Cauchy surfaces then
it must be globally isometric to de Sitter space.

Secondly, we have long felt that the  vacuum assumption
in theorem \ref{mainteo} should not be
essential, that the conclusion should still hold even
if matter is allowed a priori to be present.
In section $\ref{teomatt}$ we establish a version of theorem~\ref{mainteo} for
spacetimes satisfying the Einstein equations (with $\Lambda > 0$) with respect to a class of matter models that contains perfect fluids; see theorem \ref{matterfield}.

In the next section
we set notation, give some precise definitions and establish some preliminary results.

\section{Preliminaries}

Throughout this paper we will be using standard notation for causal
sets and relations. Refer to $\cite{P,ON}$ for the main results and
definitions in causal theory.

\subsection{Definitions and the null splitting theorem}

As usual, a spacetime $(\tilde{M},\tilde{g})$ is a connected,
time-oriented four dimensional Lorentzian manifold. Following
Penrose, we say that a spacetime $(\tilde{M},\tilde{g})$ admits a
conformal boundary $\scri$ if there exists a spacetime with
non-empty boundary $(M,g)$ such that
\begin{enumerate}
\item $\tilde{M}$ is the interior of $M$ and $\scri =\partial M$,
thus $M=\tilde{M}\cup\scri$. \item There exists $\Omega\in{\cal
C}^\infty({M})$ such that
\begin{enumerate}
\item ${g}=\Omega^2\tilde{g}$ on $\tilde{M}$, \item $\Omega >0$ on
$\tilde{M}$, \item $\Omega =0$ and $d\Omega \neq 0$ on $\scri$.
\end{enumerate}
\end{enumerate}
In this setting $g$ is referred to as the unphysical metric, $\scri$
is called the conformal boundary of $\tilde{M}$ in $M$ and $\Omega$
its defining function.

Further, we will say a spacetime $(\tilde{M},\tilde{g})$ admitting
a conformal boundary $\scri$ is \textit{asymptotically de Sitter}
if $\scri$ is spacelike. Thus, by considering the standard
conformal embedding of $dS^n$ in the Einstein cylinder we clearly
note that $dS^n$ is an asymptotically de Sitter space itself.
However, we emphasize that the definition of asymptotically de
Sitter does not require $\scri$ to be compact. This lack of
compactness causes some complications in some of the arguments.

Many physically relevant scenarios in General Relativity are
modelled by asymptotically de Sitter spacetimes. Schwarzchild de
Sitter spacetime, which  models a black hole sitting in a
positively curved background, is one such example (with a
noncompact $\scri$, in fact). Other examples can be found in the
context of cosmology, for instance the dust-filled
Friedmann-Robertson-Walker models which satisfy the Eintein
equations with $\Lambda >0$.

Because of the spacelike character of $\scri$, in an asymptotically
de Sitter spacetime, $\scri$ can be decomposed as the union of the
disjoint sets $\scri^+= \{p\in\scri\mid\nabla\Omega_p\ \textrm{is
future pointing}\}$ and $\scri^-= \{p\in\scri\mid\nabla\Omega_p\
\textrm{is past pointing}\}$. As a consequence, $\scri^+\subset
I^+(\tilde{M},M)$ and $\scri^-\subset I^-(\tilde{M},M)$. It follows
as well that both sets $\scri^+$, $\scri^-$ are acausal in $M$.

An asymptotically de Sitter spacetime is said to be
\textit{asymptotically simple}
if every inextendible null geodesic has endpoints on
$\scri$.   Such spacetimes are, in particular, null geodesically
complete.
A \emph{null line} is a globally achronal inextendible null
geodesic. Recall that a spacetime satisfying the Einstein
equations is said to obey the \textit{null energy condition} if
$T(K,K)\ge 0$ for all null vectors $K\in TM$. As theorem
$\ref{rigds}$ shows, the occurrence of a null line and the null
energy condition are incompatible for asymptotically simple
and de Sitter solutions to vacuum Einstein equations different from $dS^4$.

Theorem $\ref{rigds}$ is a consequence of the null splitting theorem
$\cite{gal01}$, which plays an important role in the proof of
theorem $\ref{mainteo}$ as well. Here is the precise statement:

\begin{teo}
Let $(M,g)$ be a null geodesically complete spacetime which obeys
the null energy condition. If $M$ admits a null line $\eta$, then
$\eta$ is contained in a smooth properly embedded, achronal and
totally geodesic null hypersurface $S$.
\end{teo}

\begin{remark}\label{remark}
The proof of the null splitting theorem actually shows how to
construct such an $S$: let $\partial_0I^{\pm}(\eta )$ be the
connected components of $\partial I^{\pm}(\eta)$ containing $\eta$,
then  $\partial_0I^+(\eta )$ and $\partial_0 I^-(\eta )$ agree and
this common surface satisfies all aforementioned properties.
Moreover, the proof also shows that future  null completeness of
$\partial_0I^-(\eta )$ and past null completeness of
$\partial_0I^+(\eta )$ are sufficient for the result to hold (see
remark IV.2 in $\cite{gal01}$.)  This point is essential to the proof
of Theorem \ref{mainteo}.
\end{remark}

\subsection{Extension lemmas}
In order to prove theorem $\ref{mainteo}$ we are faced with the
technical difficulty of dealing with a spacetime with boundary. Thus
it is convenient to think of our spacetime with boundary as embedded
in a larger open spacetime. This can always be done, as the next
result shows.

\begin{lemma}\label{openex}
Every spacetime with boundary $(M,g)$ admits an extension to a
spacetime $(N,h)$.
\end{lemma}
\par

\noindent\textit{Proof:} First extend $M$ to a smooth manifold
$M^{\prime}$ by means of attaching collars to all the components
of $\partial M$. Since $M$ is time orientable, there exists a
timelike vector field $V\in{\cal X}(M)$. Let us extend $V$ to all
of $M^{\prime}$ and let $W=\{p\in M^{\prime}\mid V_p\neq 0\}$.
Clearly $W$ is an open subset of $M^{\prime}$ containing all of
$M$, so without loss of generality we can assume
$M^{\prime}=W$.\par

Let $p\in\partial M$ and choose a $M^{\prime}$-chart ${\cal U}_p$
around it. Now let $g=g_{ij}dx^idx^j$ be the coordinate expression
of $g$ in the $M$-chart $M\cap{\cal U}_p$. Since the $g_{ij}$'s
are smooth functions on $M\cap{\cal U}_p$, they can be smoothly
extended to an $M^{\prime}$-neighborhood ${\cal
U}^{\prime}_p\subset{\cal U}_p$ with $M\cap{\cal
U}^{\prime}_p=M\cap{\cal U}_p$. Let us denote by $g^{\prime}_{ij}$
such extensions. It is important to notice that ${\cal
U}^{\prime}_p$ can be chosen in such a way that
$g^{\prime}=g^{\prime}_{ij}{dy}^i{dy}^j$ is a Lorentz metric
 on ${\cal U}^{\prime}_p$ with $g^{\prime}(V,V)<0$. Choose a cover $\{{\cal
U}_{\alpha}\}$ of $\partial M$ by such open sets and let us define
$h_{\alpha}=2e_0^*\otimes e_0^*+g^{\prime}_{\alpha}$ on $\{{\cal
U}_{\alpha}\}$, where $e_0$ denotes the unit vector field (with respect to $g^{\prime}$) in the direction of $V$. Further
consider a smooth partition of unity $f_{\alpha}$ subordinated to
$\{{\cal U}_{\alpha}\}$, thus
$h_0=\sum_{\alpha}f_{\alpha}h_{\alpha}$ is a Riemannian metric on
${\cal U}=\cup_{\alpha}{\cal U}_{\alpha}$.\par

Finally,  let $X$ be the unit vector field (with respect to $h_0$)
in the direction of $V$, let $\omega$ be the covector $h_0$-related
to $X$ and let $g^{\prime\prime}=h_0-2\omega\otimes\omega$. It is
straightforward  to check that $g^{\prime\prime}$ is a
Lorentz metric on ${\cal U}$ that agrees with $g$ on the overlap
${\cal U}\cap M$. Thus by gluing $g^{\prime\prime}$ and $g$
together we obtain a Lorentz metric $h$ on
$N={\cal U}\cup M$. Notice $h$ is smooth
since ${\cal U}$ is open. $\Box$\par

Now that we have successfully extended our spacetime with boundary,
we would like to verify that our extension inherits some important
causal properties. More precisely, we show that global
hyperbolicity extends ``beyond $\scri$" in the asymptotically de
Sitter setting. That is, if $(\tilde{M},\tilde{g})$ is globally
hyperbolic, then we can choose a globally hyperbolic extension
$(N,h)$ of it.

\begin{lemma}\label{ghex}
Let $(\tilde{M},\tilde{g})$ be a globally hyperbolic and
asymptotically de Sitter spacetime, then $(\tilde{M},\tilde{g})$
can be embedded in a globally hyperbolic spacetime $(N,h)$ such
that $\scri$ topologically separates $\tilde{M}$ and
$N-\tilde{M}$.
\end{lemma}

\noindent\textit{Proof:} It suffices to show $(\tilde{M},\tilde{g})$
can be extended past $\scri^-$ since a similar procedure can be used
to extend $(\tilde{M},\tilde{g})$ beyond $\scri^+$, thus without
loss of generality we can assume $\scri=\scri^-$. By lemma
\ref{openex} there is an open spacetime $(N_0,h)$ extending $(M,g)$.
Since $N_0$ is obtained from $M$ by attaching collars, the
separation part of the proposition holds. As a consequence $\scri$
is acausal in $N_0$, hence the Cauchy development $D(\scri , N_0)$
is an open subset of $N_0$. Thus $N=M\cup D(\scri , N_0)$ is an open
spacetime containing $M$.  We claim  that  $(N,h)$ is globally
hyperbolic. In fact, it is easy to see that if $S$ is a Cauchy
surface for $(\tilde{M},\tilde{g})$ then it is also a Cauchy surface
for $(N,h)$.    Indeed, any inextendible causal curve in $N$ must
meet $M$,  and hence will intersect~$S$. $\Box$\par

\section{Rigidity without  asymptotic simplicity}\label{teomain}

The main aim of this section is to prove the following theorem and
discuss some of its consequences:

\begin{teo}\label{mainteo1}
Let $(\tilde{M},\tilde{g})$ be a globally hyperbolic and
asymptotically de Sitter spacetime of dimension $n=4$ satisfying the
vacuum Einstein equations with positive cosmological constant. If
$\tilde{M}$ has a null line with endpoints $p\in\scri^-$,
$q\in\scri^+$ then $(\tilde{M},\tilde{g})$ is isometric to an open
subset of de Sitter space containing a Cauchy surface.
\end{teo}

\begin{figure}
\begin{center}
\includegraphics[scale=.4]{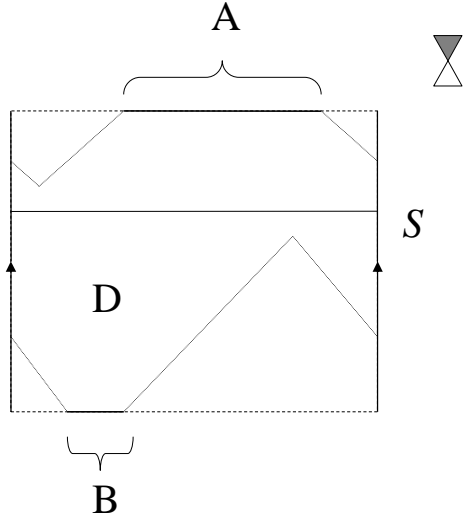} \caption{\label{fig1}
{\footnotesize $D=I^+(B)\cap I^-(A)$ is an asymptotically de
Sitter and globally hyperbolic open subset of de Sitter space with
Cauchy surface $S$.}}
\end{center}
\end{figure}

Before moving into the proof, we would like to comment that the
result is sharp, in the sense that there exists globally
hyperbolic proper subsets of $dS^4$ which contain a null line with
endpoints in $\scri$ (see fig. $\ref{fig1}$ above); see however
theorem \ref{maximal}. We remark also that some globally
hyperbolic and asymptotically de Sitter spacetimes, such as
Schwarzschild de Sitter space, do possess null lines although they
do not extend to $\scri$.

We begin the proof of theorem $\ref{mainteo1}$ by considering a
couple of technical lemmas, which establish that the achronal
boundaries $\partial I^{+}(\eta )$, $\partial I^{-}(\eta )$ are the
result of exponentiating the respective null cones about the
endpoints of $\eta$ in $\scri$.

\begin{lemma}\label{structurelemma0}
Let $(\tilde{M},\tilde{g})$ be a globally hyperbolic and
asymptotically de Sitter spacetime and $\eta$ a future directed
causal curve in $M$. Further assume $p\in\scri^-$ is the past
endpoint of $\eta$. Then
\begin{enumerate}\item $ { \partial I^+(\eta )=J^+(p,N)-(I^+(p,N)\cup\{ p\})}$,
\item $ J^+(N_p,N)\cap\tilde{M}\subset D^+(N_p,N)\cap\tilde{M}$,
\end{enumerate}
\noindent where $N_p:=\partial_NI^+(p,N)$ and $N$  is a globally
hyperbolic extension of $(M,g)$.
\end{lemma}

\noindent\textit{Proof:}  First notice that by global hyperbolicity the set
$J^+(p,N)$ is closed in $N$, and as a consequence
\begin{equation}\label{eqq01}
\partial_NI^+(p,N)=J^+(p,N)-I^+(p,N).
\end{equation}
Thus by the acausality of $\scri^-$ we have
\begin{equation}\label{eqq02}
\tilde{M}\cap\partial_NI^+(p,N)=\partial_NI^+(p,N)-\{p\}
\end{equation}
Let us show now $I^+(\eta )=I^+(p,N)$. It is clear
that $I^+(\eta )\subset I^+(p,N)$. Conversely, let $x\in
I^+(p,N)$ and let us take
$y\in\eta\cap I^-(x,N)$. Since any future timelike curve from $y$ to
$x$ has to be contained in $\tilde{M}$ due to the separating
properties of $\scri^-$, we have $x\in I^+(\eta )$ and thus
$I^+(p,N)\subset I^+(\eta )$ is proven. As a consequence $\partial
I^+(\eta )=\partial_{\tilde{M}}I^+(p,N)$. Finally, since $I^+(p,N)$ is an open set in $N$  we get
\begin{equation}\label{eqq03}
\partial_{\tilde{M}}I^+(p,N)=\tilde{M}\cap\partial_NI^+(p,N).
\end{equation}
Then the first assertion follows.

To prove the second part of the lemma we proceed by contradiction.
Thus let us assume $x\in
J^+(N_p,N)\cap\tilde{M}-D^+(N_p,N)\cap\tilde{M}$, hence it follows $x\in I^+(p,N)$. On the other hand, since  $x \notin D^+(N_p,N) \cap \tilde M$,
there is a past
inextendible causal curve $\gamma $ starting at
$x$ that does not intersect $N_p$.  Notice $\gamma$ never leaves
$I^+(p,N)$, since otherwise it had to intersect
$N_p=\partial_NI^+(p,N)$. Thus $\gamma$ is contained in the
compact set $J^+(p,N)\cap J^-(x,N)$, contradicting strong causality. $\Box$
\par

In a time dual manner if $\eta$ has a future endpoint $q\in\scri^+$,
we get $\partial I^-(\eta )=J^-(q,N)-(I^-(q,N)\cup\{ q\})$.

\begin{lemma}\label{structurelemma1}
Let $(\tilde{M},\tilde{g})$ be a globally hyperbolic and
asymptotically de Sitter spacetime and let $\eta$ be a future
directed null line in $\tilde{M}$ having endpoints $p\in\scri^-$
and $q\in\scri^+$. Further assume $(\tilde{M},\tilde{g})$
satisfies the null energy condition. Then $\partial I^+(\eta )$ is
the diffeomorphic image under the exponential map $\exp_p$ of the
set $(\Lambda_p^+-\{0_p\})\cap {\cal O}$ where $\Lambda_p^+\subset
T_pM$ is the future null cone based at $0_p$
and
${\cal O}$ is the biggest open set on which $\exp_p$ is defined.
\end{lemma}

\noindent\textit{Proof:} Let $(N,h)$ be as in the previous lemma.
Hence by  lemma $\ref{structurelemma0}$, any point in
$\tilde{M}\cap\partial_NI^+(p,N)$ is the future endpoint of a
future null geodesic segment emanating from $p$. Thus $\partial
I^+(\eta )\subset\exp_p((\Lambda_p^+-\{0_p\})\cap {\cal
O})\cap\tilde{M}$.

Now let $\gamma$ be a null generator of $\partial I^+(\eta )$
passing through $x\in \partial I^+(\eta )$. Let $y\in\gamma$ a
point slightly to the past of $x$ and notice
$y\in\partial_NI^+(p,N)$ by equation $(\ref{eqq02})$. On the other
hand, let $\overline{\gamma}(t)$ be a null geodesic emanating from
$p$ and passing through $y$. Then $\gamma$ coincides with
$\overline{\gamma}\subset\tilde{M}$ since otherwise we would have
two null geodesics meeting at an angle in $y$ and hence $x\in
I^+(p,N)$.
Thus, $\gamma$ can be extended to $p\in\scri^-$ and thus it is
past complete. In a time dual fashion, the generators of $\partial
I^-(\eta )$ are future complete.

Let $S$ be the component of $\partial I^+(\eta )$ containing $\eta$.
By the proof of the null splitting theorem, $S$ is a closed smooth
totally geodesic null hypersurface in $\tilde{M}$. (Here we are
using the fact that the null splitting theorem does not require full
null completeness; see remark \ref{remark}.) As a consequence, the
null generators of $S$ do not have future endpoints in $\tilde{M}$
and hence are future inextendible in $S$. Furthermore, by the
argument in the previous paragraph, each of these generators is the
image under $\exp_p$ of the set $\mathbf{V}\cap{\cal O}$, where
$\mathbf{V}$ is an inextendible null ray in $\Lambda^+_p$.

Let $\gamma$ be a generator of $S$, then $\gamma\cap
I^+(p,N)=\emptyset$. Thus $\gamma$ is conjugate point free and
does not intersect with any other generator of $S$. As a result we
have that $S$ is the diffeomorphic image under $\exp_p$ of an open
subset of~$\Lambda_p^+-\{0_p\}$.

To check that $S$ encompasses the whole local future null cone at
$p$, let us consider a causally convex normal neighborhood ${\cal
V}$ of $p$ and a spacelike hypersurface $\Sigma$ slightly to the
future of $\scri^-$. Thus $\Sigma_0:=\Sigma\cap
\exp_p((\Lambda_p^+-\{0_p\})\cap {\cal V})$ is connected.
Moreover, by the way ${\cal V}$ and $\Sigma$ were chosen we have
$\Sigma_0\subset J^+(p)-(I^+(p)\cup \{p\})=\partial I^+(\eta )$.
Thus $\exp_p((\Lambda_p^+-\{0_p\})\cap {\cal
O})\cap\tilde{M}\subset S$ since every future null geodesic
emanating from $p$, including $\eta$, must intersect $\Sigma_0$.
It follows $S=\partial I^+(\eta )$ and the proof is complete.
$\Box$\par

\smallskip
Now we start the proof of the main result of this section.

\noindent\textit{Proof of theorem $\ref{mainteo1}$:} We first show
that $(\tilde{M},\tilde{g})$ has simply connected Cauchy surfaces.
To this end, let $\partial_0I^+(\eta)$, $\partial_0I^-(\eta )$ be
the components of $\partial I^+(\eta )$, $\partial I^-(\eta )$
containing $\eta$ respectively. By the the null splitting theorem,
we have $\partial_0I^+(\eta)=\partial_0I^-(\eta  )$, and this common
null hypersurface is closed, smooth and totally geodesic. Moreover,
by the previous lemma we also conclude $S:=\partial I^+(\eta )$ is
connected, i.e. $S=\partial_0 I^+(\eta )$. Lastly, by lemma
$\ref{structurelemma0}$ we have $\partial I^+(\eta )=N_p-\{p\}$ and
$\partial I^-(\eta )=N_q-\{q\}$. Thus $N_p-\{p\}=S=N_q-\{q\}$. On
the other hand, notice that the equality, $N_p-\{p\}=N_q-\{q\}$, in
conjunction with lemma $\ref{structurelemma1}$, imply that every
point in $S$ is at the same time the future endpoint of a null
geodesic emanating from $p$ and the past endpoint of a null geodesic
from $q$. These geodesic segments must form a single geodesic,
otherwise achronality of $\eta$ would be violated. Hence, all future
null geodesics emanating from $p$ meet again at $q$. Then $\calS =
S\cup\{p,q\}$ is homeomorphic to a sphere.  By a suitable small
deformation of $\calS$ near $p$ and $q$, we obtain an achronal
hypersurface $\calS'$ in $\tilde M$ homeomorphic to an
($n-1$)-sphere.  Using the compactness of $\calS'$ and basic
properties of Cauchy horizons, one easily obtains,
 $H^-(\calS') = H^+(\calS') = \emptyset$, and hence $\calS'$ is a Cauchy surface
 for $\tilde M$.

As our next step, we proceed to show $(\tilde{M},\tilde{g})$ has
constant curvature. Let $(N,h)$ be a globally hyperbolic extension
of $(M,g)$, then by lemma $\ref{structurelemma0}$ we have
${\displaystyle I^+(S)\subset D^+(N_p,N)\cap \tilde{M}}$. In a time
dual fashion ${\displaystyle I^-(S)\subset D^-(N_q,N)\cap
\tilde{M}}$, hence as a consequence of proposition $[3.15]$ in
$\cite{P})$ we get $\tilde{M}=I^+(S)\cup S\cup I^-(S)$. Thus
$\tilde{M}\subset D^+(N_p,N)\cup D^-(N_q,N)$.

Now recall that $S$ is a totally geodesic null hypersurface. As a
consequence the shear tensor $\tilde{\sigma}_{\alpha\beta}$ of $S$
in the physical metric $\tilde{g}$ vanishes, and since the shear
scalar $\tilde{\sigma}=
\tilde{\sigma}_{\alpha\beta}\tilde{\sigma}^{\alpha\beta}$ is a
conformal invariant we have $\sigma_{\alpha\beta}\equiv 0$ as
well. Then from the propagation equations (cfr. $[4.36]$ in
$\cite{HE}$) we deduce that the components $W_{\alpha 0 \beta 0}$
of the Weyl tensor vanishes on $S$, where $\{e_0,e_1,e_2,e_3\}$ is
a null tetrad with $e_0$ adapted to the null generators of $S$. In
$\cite{F}$, Friedrich used the conformal field equations
\begin{equation}
\nabla_{\alpha}{d^{\alpha}}_{\beta\gamma\zeta}=0,\qquad
{d^{\alpha}}_{\beta\gamma\zeta}=\Omega^{-1}{W^{\alpha}}_{\beta\gamma\zeta}
\end{equation}
along with a recursive ODE argument to guarantee the vanishing of
the rescaled conformal tensor $d$ on $D^+(S\cup\{p\},N)$ given that
$W_{0000}$ vanishes on $S$. Hence, we have shown $d\equiv 0$ on
$D^+(N_p,N)$. Thus by the conformal invariance of the Weyl tensor we
have $\widetilde{W}\equiv 0$ on $D^+(N_p,N)\cap\tilde{M}$. By a time
dual argument we conclude $\widetilde{W}\equiv 0$ on
$D^-(N_q,N)\cap\tilde{M}$, thus $\widetilde{W}\equiv 0$ on
$\tilde{M}$. Finally, since $(\tilde{M},\tilde{g})$ satisfies the
vacuum Einstein equations with positive cosmological, the vanishing
of the Weyl tensor implies that $(\tilde{M},\tilde{g})$ has constant
curvature $C>0$. Note that this is the only part of the argument
where the hypothesis $n=4$ is used.

Further, since $(\tilde{M},\tilde{g})$ is simply connected, there
exists a local isometry $\Phi\colon\tilde{M}\to dS^4$ by the
Cartan-Ambrose-Hicks Theorem $\cite{CE,ON}$.
(However, since $(\tilde{M},\tilde{g})$
needn't be complete, $\Phi$ needn't be a covering map.)

Then the theorem follows
by a direct application of the following result.~$\Box$\par

\begin{prop}\label{embed}
Let $(\tilde{M},\tilde{g})$ be a globally hyperbolic
spacetime with compact Cauchy surfaces. If there exists a local
isometry $\Phi\colon \tilde{M}\to dS^n$, then
$(\tilde{M},\tilde{g})$ is isometric to an open subset of $dS^n$
containing a Cauchy surface.
\end{prop}

\noindent\textit{Proof:} We need to show that $\Phi$ is
injective.   Let us denote by ${{\cal S}}$ a fixed
Cauchy surface of $\tilde{M}$.  By virtue of $\cite{BS}$, we can
assume that ${\cal S}$ is  smooth and spacelike,
and in fact that $\tilde{M}=\R \times{\cal S}$,
with each slice $\calS_a = \{a \} \times \calS$ a smooth compact spacelike
Cauchy surface.
We proceed to show  that $\Phi_{\cal S} := \Phi \circ i : \calS \to
dS^4$  ($i =$  inclusion) is an
 embedding. To this end,  let $\mathfrak{ S}$ be a fixed Cauchy
surface for $dS^4$, and let $\pi : dS^4 \to \mathfrak{ S}$ be projection
along the integral curves of a timelike vector field on $dS^4$ into
$\mathfrak{ S}$. Further, let $\hat{\cal S}:=\Phi ({\cal S})$.

We first show $\pi\vert_{\hat{\cal S}}$ is a local homeomorphism.
Since $\hat\calS$ is compact,
%manifolds are locally compact Hausdorff spaces,
it suffices to show $\pi$ is locally one to one. Thus let $y\in\hat{\cal S}$.
Take then $x\in {\cal S}$ with $\Phi (x)=y$ and consider a
neighborhood ${\cal V}$ of $x$ such that $\Phi\vert_{{\cal V}}$ is
an isometry. Further, since $dS^n$ is globally hyperbolic there is
a causally convex neighborhood ${\cal U}$ of $y$ contained in
$\Phi ({\cal V})$. Let then $a,b\in {\cal U}$ such that $\pi
(a)=z=\pi (b)$. If $a\neq b$ let us denote by $\gamma$ the portion
of $\pi^{-1}(z)$ from $a$ to $b$, then $\gamma$ is a timelike
curve connecting $a$ and $b$. Thus by causal convexity, $\gamma$
must be contained in ${\cal U}\subset\Phi ({\cal V})$. Hence
$\Phi^{-1}(\gamma )\cap {\cal V}$ is a timelike curve joining two
points of ${{\cal S}}$. But ${{\cal S}}$ is achronal, being a
Cauchy surface for $\tilde{M}$. Thus $a=b$ so $\pi\vert_{\hat{\cal
S}\cap{\cal U}}$ is injective.

Hence $F\colon {\cal S}\to\mathfrak{S}$ defined by
$F=\pi\circ\Phi_{{\cal S}}$ is a local homeomorphism.  Further,
since $\cal S$ is compact, $F$ is proper.
Thus by a standard
topological result (refer for instance to
 proposition 2.19 in
 $\cite{LEES}$ and notice that the proof works as well in the continuous setting)
 we have that $F$ is a topological covering map.
Moreover, since $\mathfrak{S}$ is simply connected we have that
$F$ is injective, hence a homeomorphism. Thus $\Phi_{{\cal S}}$ is
injective as well, therefore a smooth embedding since ${\cal
S}$ is compact.

Then $\hat{{\cal S}}$ is a compact embedded spacelike hypersurface
in $dS^n$.   But a compact spacelike hypersurface in a globally hyperbolic
spacetime is necessarily Cauchy (cfr. \cite{budic}).
Thus,  $\hat{{\calS}}$ is a Cauchy surface, and in particular  is achronal.
Clearly the same conclusion applies to $\hat{{\cal S}}_a:=\Phi ({{\cal S}_a})$
for each $a \in \R$.
Since $\hat{{\cal S}}_a:=\Phi ({{\cal S}_a})$ is
achronal for all $a\in\R$ it follows that no two of these surfaces
can intersect. Thus $\Phi$ is injective.

The result now follows since every injective local isometry is an
isometry onto an open subset of the codomain. $\Box$

\begin{remark}
G. Mess points out in $\cite{messy}$ the existence of  simply
connected and locally de Sitter spacetimes (i.e., spacetimes of constant curvature
$\equiv 1$) that can not be isometrically embedded into 3-dimensional de Sitter
space. In $\cite{beng}$, Bengtsson and Holst were able to
construct a similar example in dimension four. Moreover, this latter
spacetime occurs as a Cauchy development of a Cauchy surface $S$
with  noncompact topology $\mathbb{H}^2\times\R$.
On the other hand, proposition \ref{embed}  shows that no such
example can be found having compact Cauchy surfaces.
\end{remark}

We end this section by noting that if a spacetime satisfies all
hypotheses of theorem $\ref{mainteo1}$ and arises as the
evolution of Cauchy data, it is isometric to $dS^4$. Recall the
fundamental result by Choquet-Bruhat and Geroch
$\cite{YCBG}$ that establishes the existence of a maximal Cauchy
development ${\cal M}^*$ relative to a initial data set $(S,h,K)$
satisfying the vacuum Einstein equation. Moreover, such a set
satisfies a domain of dependence condition $\cite{YCBG, Wald}$:

\begin{teo}\label{domdep}
Let $(S_i,h_i,K_i)$, $i=1,2$, be two initial data sets with
maximal Cauchy developments $({\cal M}^*_i,g_i^*)$. Let
$A_i\subset S_i$ and assume there is a diffeomorphism sending
$(A_1,h_1,K_1)$ to $(A_2,h_2,K_2)$. Then $D(A_1,{\cal M}^*_1)$ is
isometric to $D(A_2,{\cal M}^*_2)$.
\end{teo}

As pointed out in $\cite{larsgal}$, the argument used in
$\cite{YCBG}$ is also valid when considering the Einstein
equations with cosmological constant. Thus we have:

\begin{teo}\label{maximal}
Let $(S,h,K)$ be an initial data set and $({\cal M}^*,g^*)$ its
maximal Cauchy development. Suppose $({\cal M}^*,g^*)$ is
asymptotically de Sitter and satisfies the vacuum Einstein
equations.  If $({\cal M}^*,g^*)$ contains a null line from
$\scri^-$ to $\scri^+$, then it is isometric to $dS^4$.
\end{teo}

\noindent\textit{Proof:} By theorem $\ref{mainteo1}$ there is an
isometry $\Phi\colon({\cal M}^*,g^*)\to {\cal A}$, where ${\cal
A}$ is an open subset of $dS^4$. Furthermore, by the proof of
theorem $\ref{mainteo1}$ we also know $\Phi (S)$ is a Cauchy
surface of $dS^4$, hence $D(\Phi (S),dS^4)=dS^4$. Then the result
follows from theorem $\ref{domdep}$. $\Box$\Par

\section{The non-vacuum case}\label{teomatt}

In this section we generalize theorem $\ref{mainteo1}$ to spacetimes
satisfying the Einstein equations
\begin{equation}
\text{Ric}-\frac{1}{2}Rg+\Lambda g=T
\end{equation}
where the energy momentum tensor $T$ is that of matter. More
specifically, we will be considering matter fields on an
asymptotically de Sitter spacetime $(\tilde{M},\tilde{g})$
satisfying all four of the following hypotheses, which are
satisfied by perfect fluids:

\medskip
\noindent \textbf{A. The Dominant Energy Condition.}

Recall that $T$ satisfies the Dominant Energy Condition if for all
timelike $X\in{\cal X}(M)$, $T(X,X)\ge 0$ and the vector field
metrically related to $T(X,-)$ is causal. It is easy to see that a
perfect fluid satisfies the dominant energy condition if and only if
$\rho \ge\vert p\vert $.

\medskip
\noindent\textbf{B. $\widetilde{\text {Tr}}\, T\le 0$ on a
neighborhood of $\scri$.}

This hypothesis is satisfied for a wide verity of fields. It holds
for  photon gases,  electromagnetic fields $\cite{SW,kriele,Wald}$
as well as for quasi-gases $\cite{SW}$. In particular it holds for
dust, pure radiation and all perfect fluids satisfying $0\le p\le
\rho /3$.

\medskip
\noindent\textbf{C. If $K$ is a null vector at $p\in \tilde{M}$
with $T(K,K)=0$, then $T\equiv 0$ at $p$.}

Recall that a Type I energy-momentum tensor is by definition
diagonalizable $\cite{HE}$. With the exception of a null fluid,
all energy-momentum tensors representing reasonable matter are
diagonalizible $\cite{Wald}$. Let $\{\rho ,p_1,p_2 ,p_{3}\}$ be
the eigenvalues of such a tensor with respect to an orthonormal
basis $\{e_0,e_1,e_2 ,e_3\}$, where $e_0$ is timelike. Then for a
Type I tensor the existence of $\lambda\in (0,1)$ satisfying
$\lambda \rho\ge\vert p_i\vert $, $i=1,2,3$ prevents the vanishing
of $T_x$ in null directions, unless $T_x\equiv 0$. In particular,
perfect fluids with $0\le p\le \rho /3$ satisfy this condition.

\medskip
\noindent\textbf{D. The  following fall-off condition holds:}
\begin{equation}\label{falloff2}
\displaystyle{\lim_{x\to\scri}\Omega {T(\nabla \Omega ,\nabla\Omega
)}=0}.
\end{equation}

For instance, for $4$-dimensional dust-filled FRW models
with $\Lambda
> 0$, we have $\Omega T(\nabla \Omega,\nabla \Omega )\sim \rho
/\Omega$ near $\scri$, whereas $\rho \sim \Omega^3$, so that
\eqref{falloff2} is easily satisfied.  A similar conclusion holds
for more general perfect fluids with suitable equation of state.

\begin{teo}\label{matterfield} Let $(\tilde{M}, \tilde{g})$ be a globally hyperbolic and
asymptotically de Sitter spacetime  which is a solution of the
Einstein equations with positive cosmological constant
\begin{equation}\label{einstein}
R_{\alpha\beta}-\frac{1}{2}Rg_{\alpha\beta}+\Lambda
g_{\alpha\beta}=T_{\alpha\beta},
\end{equation}
 where the
energy-momentum tensor $T$ satisfies conditions \textbf{A} -
\textbf{D} above. If $(\tilde{M}, \tilde{g})$ contains a null line
$\eta$ with endpoints on $\scri$ then $(\tilde{M}, \tilde{g})$ is
isometric to an open subset of de Sitter space containing a Cauchy
surface.
\end{teo}

\noindent\textit{Proof:}  The goal is to show that the energy-momentum tensor
$T$ vanishes on $\tilde M$, so that theorem \ref{matterfield} reduces to
theorem \ref{mainteo}.  We begin by showing that after a suitable
gauge fixing, the unphysical metric assumes a convenient form near
$\scri^-$ (and time-dually, near $\scri^+$).

\begin{lemma}\label{gauge}
Let $(\tilde{M}, \tilde{g})$ be as in theorem \ref{matterfield}.
Then $\Omega$ and $g$ can be chosen so that  in  a neighborhood $\mathcal{U}$ of $\scri^-$, $\Omega$ measures distance to $\scri^-$ with respect to $g$, and
$\tilde g$ takes the form,
\begin{equation}\label{gauss}
\tilde{g}=\frac{1}{\Omega^2}[-d\Omega^2+h(u)]\qquad
\text{on ${\cal U}$}  \,,
\end{equation}
where $h(u)$ is a Riemannian metric on the slice
$S_u=\Omega^{-1}(u)$.
Moreover, these choices can be made so that the fall-off
condition {\bf D} still holds.
\end{lemma}

\smallskip
\noindent
{\it Proof of the lemma:}
Following a computation in
$\cite{larsgal}$ we note that the fall-off condition  $\textbf{D}$
implies that
\beq\label{unit}
g(\nabla \Omega ,\nabla\Omega )= -1 \qquad \text{on $\scri^-$}  \,.
\eeq

Consider now the conformally rescaled quantities
$\overline{\Omega}={\Omega}/{\theta}$,
$\overline{g}={g}/{{\theta}^2}$; then we want to find $\theta$
smooth in a neighborhood ${\cal U}$ of $\scri^-$ such that
$\overline{\Omega}$ agrees with $\Omega$ on $\scri^-$ and
$\overline{g}(\overline{\nabla}\,\overline{\Omega},\overline{\nabla}\,\overline{\Omega})=-1$ on ${\cal U}$. To do so, we notice that this latter equation
gives rise to the first order PDE
\begin{equation}\label{011}
2\theta g(\nabla\Omega ,\nabla\theta )-\Omega g(\nabla\theta,\nabla\theta ) -
\theta^2 a = 0  \,,
\end{equation}
where by \eqref{unit},
$a:=
\Omega^{-1}(1+g(\nabla\Omega,\nabla\Omega ))$ is smooth. By a
standard PDE result (refer to the generalization of theorem 10.3
on page 36 in $\cite{Spivak5}$) this equation subject to the
initial condition $\theta{\vert}_{\scri^-}=1$ has a unique
solution in a neighborhood ${\cal U}$ of $\scri^-$. Notice that,
by shrinking ${\cal U}$ if necessary, we can extend $\theta$
smoothly to a positive function in all of $M$. Since the  integral
curves of the gradient $\overline{\nabla}\, \overline{\Omega}$ are
unit speed timelike geodesics in ${\cal U}$ normal to $\scri^-$,
by further restricting ${\cal U}$ to a normal neighborhood of
$\scri^-$, we can take the slices $S_u$ to be the normal gaussian
foliation of ${\cal U}$ with respect to $\scri^-$. Thus we have
\begin{equation}\label{013}
\tilde{g}=\frac{1}{{\overline{\Omega}}^2}[-d{\overline{\Omega}}^2+h(u)]\qquad\text{on
${\cal U}$}
\end{equation}
where $h(u)$ is a Riemannian metric on the slice
$S_u=\overline{\Omega }^{-1}(u)$. Finally, notice that
\begin{equation}
T(\overline{\nabla}\, \overline{\Omega}, \overline{\nabla}\,
\overline{\Omega})=\theta^2T(\nabla\Omega ,\nabla\Omega ) +
O(\Omega )\qquad\text{on ${\cal U}$}
\end{equation}
hence the fall-off condition \textbf{D} holds for
$\overline{\nabla}\, \overline{\Omega}$ as well.
This completes the proof of the lemma.

\smallskip
Henceforth, we assume $\Omega$, $g$ have been chosen in accordance
with Lemma~\ref{gauge}.

\begin{figure}
\begin{center}
\includegraphics[angle=0,scale=.6]{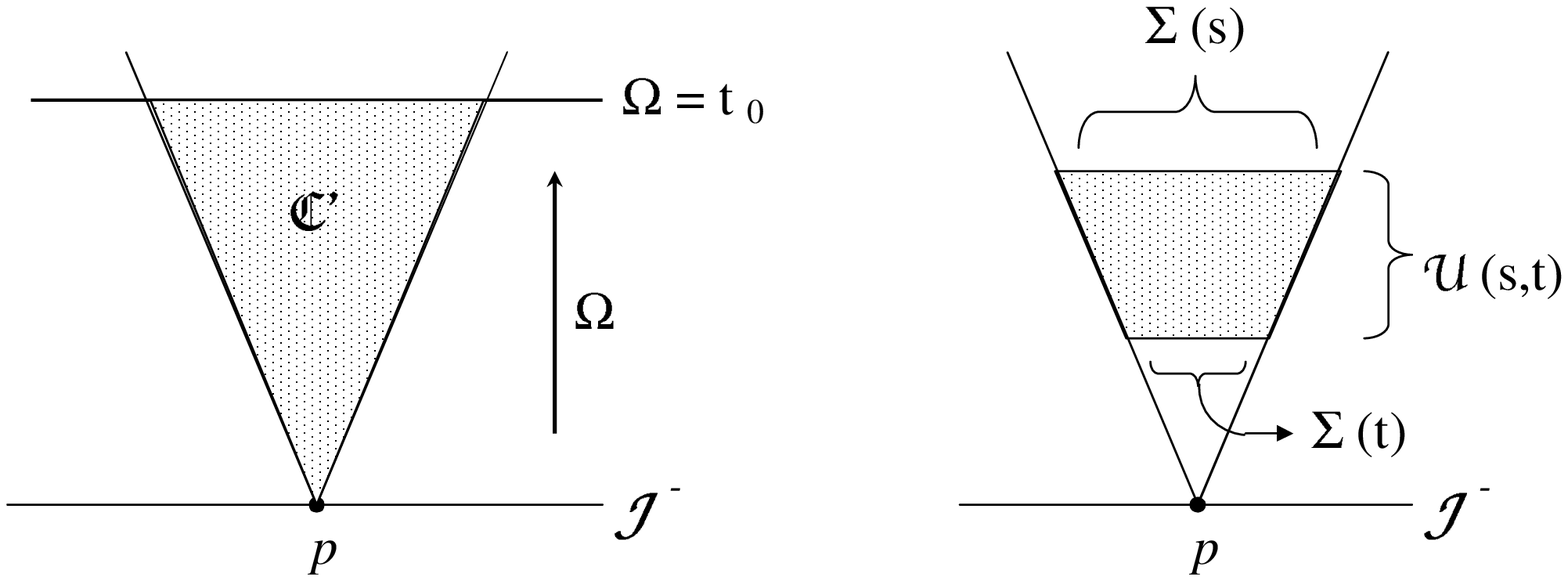} \caption{\label{fig2}}
\end{center}
\end{figure}

Recall that by lemma $\ref{structurelemma1}$ the set ${\cal
S}:=\partial I^+(\eta )$ is just the future null cone at $p$, i.e.
${\cal S}=\exp_p(\Lambda^+_p\cap{\cal O})\cap\tilde{M}$ where ${\cal
O}$ is the maximal set in which $\exp_p$ is defined. Let us denote
now the local causal cone at $p$ by
$\mathfrak{C}:=\exp_p(C^+_p\cap{\cal O})\cap\tilde{M}$, hence
$\mathfrak{C}-\{p\}$ is a manifold-with-boundary and
$\partial(\mathfrak{C}-\{p\})={\cal S}$. Further let $t_0>0$ be such
that
$\mathfrak{C}^{\prime}:=\mathfrak{C}\cap\Omega^{-1}([0,t_0])\subset
{\cal U}$. For  $s,t\in (0,t_0)$ with $s<t$ we define ${\cal
U}(s,t):={\mathfrak{C}}^{\prime}\cap\Omega^{-1}([s,t])$, ${\cal
S}(s,t):={\cal S}\cap\Omega^{-1}([s,t])$ and $\Sigma
(t)={\mathfrak{C}}^{\prime}\cap\Omega^{-1}(t)$. (See figure
$\ref{fig2}$.) Thus ${\cal U}(s,t)$ is a compact manifold with
corners and $\partial {\cal U}(s,t)={\cal S}(s,t)\cup\Sigma
(s)\cup\Sigma (t)$.

The following claim is the heart of the proof of Theorem \ref{matterfield}.

\medskip
\noindent
{\bf Claim.} {\it The energy-momentum tensor $T$ vanishes on
$\mathfrak{C}^{\prime}$.}

\medskip

\noindent
{\it Proof of the claim:}
For the time being, let $s\in (0,t_0)$ be fixed and let  ${\cal
U}(t):={\cal U}(s,t)$, ${\cal S}(t):={\cal S}(s,t)$ for all $t\in
(s,t_0)$. Let $A$ be the vector field defined by $g(A, X)=T(\nabla \Omega ,X)$ for
all $X\in{\cal X}(\tilde{M})$,  hence by Stokes theorem
\begin{equation}
 \int\limits_{{\cal U}(t)}{\rm div}A\, \text
{dv}=\int\limits_{\partial{\cal U}(t)}i_{A}\text{dv}=\int\limits_{\Sigma (s )}i_{A}\text{ dv}+\int\limits_{\Sigma
(t)}i_{A}\text{dv}+\int\limits_{{\cal S}(t)}i_{A}\text{dv}.
\end{equation}

We proceed to show the integral over the null cone portion ${\cal
S}(t)$ vanishes. Thus let $x\in S$. By virtue of assumption
$\textbf{C}$, it suffices to show that~$T(K,K)=0$ for some null
vector $K\in T_x\tilde{M}$. Hence let us consider a future null
generator $\gamma$ of $S$ through $x$.
By the Raychaudhuri equation, we have
\begin{equation}
\frac{d\theta}{ ds}=-\textrm
{Ric}({\gamma}^{\prime},{\gamma}^{\prime}) -{\sigma}^2-\frac{1}
{2}{\theta}^2 \,,
\end{equation}
where $\theta$ is the null expansion (or null mean curvature) of $S$.
Since $S$ is totally geodesic by lemma $\ref{structurelemma1}$ we
must have $\theta \equiv 0$ and $\sigma\equiv 0$, thus
$\textrm{Ric}(\gamma^{\prime},\gamma^{\prime})=0$. Further, since
$\gamma^{\prime}$ is null the Einstein equations imply
$\textrm{Ric}(\gamma^{\prime},\gamma^{\prime})=T(\gamma^{\prime},\gamma^{\prime})$,
and thus $T(\gamma^{\prime},\gamma^{\prime})=0$. Hence
$i_A\text{dv}\vert_S\equiv 0$ as desired. Thus we have

\begin{equation}\label{integrales}
\int\limits_{{\cal U}(t)}\text{ div}A  \,
\text{dv}=\int\limits_{\Sigma (t )}T(\nabla\Omega ,\nabla\Omega
)\text{d}\sigma -\int\limits_{\Sigma (s )}T(\nabla\Omega
,\nabla\Omega )\text{d}\sigma .
\end{equation}

Now let $\hat{T}$ be the $(1,1)$ tensor $g$-equivalent to $T$ and
let $C$ denote tensor contraction with respect to $g$. Since
$A=C(\hat{T}\otimes\nabla \Omega)$ we have $\text{div}A =
\text{div} T\, (\nabla \Omega ) +C^2(\hat{T}\otimes \nabla (\nabla
\Omega ))$. Hence
\begin{eqnarray}\label{integr02} \int\limits_{{\cal U}(t)} \text{div} T\,  (\nabla \Omega )
\text{dv}\ \ &+&\int\limits_{{\cal U}(t)}C^2 (\hat{T}\otimes
\nabla (\nabla \Omega ) ) \text{dv}\nonumber \\
&=& \int\limits_{\Sigma (t)}\hskip -.2cm T(\nabla \Omega , \nabla
\Omega ) {\rm d}\sigma \, \text{dv} \ -\int\limits_{\Sigma
(s)}\hskip -.2cm T(\nabla \Omega , \nabla
\Omega ) {\rm d}\sigma\nonumber \\   \,.
\end{eqnarray}

Since $\mathfrak{C}^{\prime}$ is compact,  the components
${\Omega_{;\alpha}}^{;\beta}$ of $\nabla (\nabla \Omega )$ in any
$g$-orthonormal frame field are bounded from above, say by $Q$.
Similarly, $T(\nabla\Omega ,\nabla\Omega )\ge \vert
{T^{\alpha}}_{\beta}\vert$ on $\tilde{M}$ by the dominant energy
condition, hence by continuity,
$
%{\displaystyle
\lim_{z\to p}T(\nabla\Omega ,\nabla\Omega )_z \ge \lim_{z\to
p}\vert {T^{\alpha}}_{\beta}(z)\vert$ as well. Then
$C^2(\hat{T}\otimes \nabla (\nabla \Omega ) )\le  P\,
T(\nabla\Omega ,\nabla \Omega )$ on $\mathfrak{C}^{\prime}$, where
$P:=16Q$. Thus
\begin{equation}\label{integr03}
\int\limits_{{\cal U}(t)}C^2(\hat{T}\otimes \nabla (\nabla \Omega
) )\text{dv}\ \ \le \int\limits_{{\cal U}(t)}P\, T(\nabla \Omega
,\nabla \Omega ) \text{dv} \,.
\end{equation}

On the other hand, the formula relating the divergence operator of two conformally related metrics $g=\Omega^2\tilde{g}$ in a Lorentzian manifold of dimension $n$  gives,
\begin{equation}\label{cuentas06}
\text{div}\, T(\nabla \Omega )=\frac{1}{{\Omega
}^2}\widetilde{\text{div}}\, T(\nabla \Omega ) +\frac{n-2}{\Omega
}T(\nabla \Omega ,\nabla \Omega )+\frac{1}{
\Omega^3}\widetilde{\text{Tr}}\, T.
\end{equation}
Since the physical metric satisfies the Einstein equations, the
energy-momentum tensor is divergence free. Thus $\widetilde{\text{div}}\, T(\nabla \Omega )\equiv 0$ in $\tilde{M}$. Moreover, by assumption $\textbf{B}$, $\widetilde{\text{Tr}}\ T\le
0$, thus we deduce the inequality
\begin{equation}\label{enineq3}
\int\limits_{{\cal U}(t)} \text{div} T\,  (\nabla \Omega )
\text{dv}\ \le \int\limits_{{\cal U}(t)}\frac{2}{\Omega}\,
T(\nabla \Omega ,\nabla \Omega ) \text{dv}
\end{equation}

Hence equation $(\ref{integr02})$ along with $(\ref{integr03})$
and $(\ref{enineq3})$ yield
\begin{equation}\label{enineq2}
\int\limits_{\Sigma (t)}\hskip -.2cm T(\nabla \Omega ,\nabla
\Omega ) {\rm d}\sigma -\int\limits_{\Sigma (s )}\hskip -.2cm
T(\nabla \Omega ,\nabla \Omega ) {\rm d}\sigma \le
\int_{s}^t\int\limits_{\Sigma (\tau ) } \left(
\frac{2}{\Omega}+P\right) \, T(\nabla \Omega ,\nabla \Omega )
{\rm d}\sigma\,\, {\rm d}\tau.
\end{equation}

Now, we would like to analyze the limit of both sides of relation
($\ref{enineq2}$) as $s\to 0$. Let then $p(s)\in\Sigma (s)$ be
such that $T(\nabla\Omega_z ,\nabla\Omega_z )\le
T(\nabla\Omega_{p(s)},\nabla\Omega_{p(s)})$ for all $z\in\Sigma
(s)$. Such $p(s)$ always exists since $\Sigma (s)$ is compact.
Thus
\begin{eqnarray}\label{enineq4}
 \int\limits_{\Sigma (s )}
\frac{1}{\Omega }T(\nabla \Omega ,\nabla \Omega )  {\rm d}\sigma
&\le& \frac{1}{ s } T({\nabla \Omega}_{p(s )},{\nabla \Omega
}_{p(s )})\int\limits_{\Sigma (s ) }\text{d}\sigma\nonumber\\ &=&
\frac{1}{s}T({\nabla \Omega}_{p(s )},{\nabla \Omega }_{p(s
)})\text{Vol}(\Sigma (s))
\end{eqnarray}

Let us consider now a small normal neighborhood ${\cal N}$ around
$p$. It is known $\cite{volsr}$ that the metric volume of the
local causal cone truncated by a timelike vector  is of the
same order as the volume of the corresponding truncated cone in
$T_pM$.  Hence by considering $s$ very small
 we get the estimate
\begin{equation}\label{volume}
\text{Vol}(\Sigma (s))= O(s^{3}).
\end{equation}
Thus without loss of generality we can take $t_0>0$ such that
$\mathfrak{C}^{\prime}$ is contained in such a normal neighborhood
${\cal N}$.   Thus, for $s$ sufficiently small,  \eqref{enineq4}
and ($\ref{volume}$) imply,
\begin{equation}\label{enineq7}
 \int\limits_{\Sigma (s )}
\frac{1}{\Omega }T(\nabla \Omega ,\nabla \Omega )  {\rm d}\sigma
 \le C\, T({\nabla \Omega}_{p(s
)},{\nabla \Omega }_{p(s )})s^{2}
\end{equation}
for some positive constant $C$. Hence
\begin{equation}\label{limit}
\lim_{s \to 0^{+}}\int\limits_{\Sigma (s ) } \frac{1}{\Omega }T(
\nabla\Omega ,\nabla \Omega )  {\rm d}\sigma =0
\end{equation}
by virtue of assumption \textbf{D}.

Let $x=x(t)$ be the function defined by,
\begin{equation}\label{xdef}
x(t): =  \int_{0}^t\int\limits_{\Sigma (\tau ) }
\left(\frac{2}{ \Omega }+P\right)T(\nabla \Omega ,\nabla \Omega
) {\rm d}\sigma\,\, {\rm d}\tau  \,,
\end{equation}
which makes sense since, by \eqref{limit}, the integrand continuously extends to
$\tau = 0$.
By letting $s\to 0^+$ in
inequality ($\ref{enineq2}$) we obtain,
\begin{equation}\label{integr05}
\int\limits_{\Sigma (t)}\hskip -.2cm T(\nabla \Omega ,\nabla
\Omega ) {\rm d}\sigma \le x(t)  \,.
\end{equation}
Differentiation of \eqref{xdef} for $t \in (0,t_0)$ gives, \beq
\frac{dx}{dt} = \left(\frac2{t} + P\right) \int\limits_{\Sigma
(t)}\hskip -.2cm T(\nabla \Omega ,\nabla \Omega ) {\rm d}\sigma
\eeq which when combined with ($\ref{integr05}$) yields the
differential inequality,
\begin{equation}
\frac{d}{dt}\left( \frac{e^{-Pt}}{t^{2}}x \right) \le 0   \,.
\end{equation}
Hence the function
\begin{equation}
I(t)=\frac{x(t)e^{-Pt}}{t^{2}}
\end{equation}
is decreasing near $\scri^-$.

Thus, we analyze ${\displaystyle\lim_{t\to 0^+}I(t)}$. Notice
first that estimate ($\ref{enineq7}$) yields
\begin{equation}
\int\limits_{\Sigma (t ) } \left(\frac{2}{\Omega
}+P\right)T(\nabla \Omega ,\nabla \Omega ){\rm d}\sigma \le
C^{\prime}\, T(\nabla\Omega_{p(t)},\nabla\Omega_{p(t)})t^{2}
\end{equation}
for some constant $C^{\prime}\ge 0$. Thus we get
\begin{eqnarray}
\lim_{t \to 0^{+}}\frac{x(t )} {{t}^{2}} &=& \lim_{t \to
0^{+}}\frac{1}{ 2{t}}\int\limits_{\Sigma (t ) }
\left(\frac{2}{\Omega }+P\right)T(\nabla \Omega ,\nabla \Omega )
{\rm
d}\sigma\\
&\le&\frac{C^{\prime}}{2} \, t\,
T(\nabla\Omega_{p(t)},\nabla\Omega_{p(t)})\nonumber   \,,
\end{eqnarray}
which, by condition {\bf D} implies that ${\displaystyle \lim_{t \to 0^{+}}\frac{x(t )} {{t}^{2}}}=0$,
and hence ${\displaystyle\lim_{t\to 0^+}I(t)}=0$. It follows that
$I(t)\equiv 0$ on $\mathfrak{C}^{\prime}$, and consequently
$T(\nabla \Omega ,\nabla\Omega )\equiv 0$ on
$\mathfrak{C}^{\prime}$. Therefore $T\equiv 0$ on
$\mathfrak{C}^{\prime}$ by the dominant energy condition.
This completes the proof of the claim.

\smallskip
Now let $0<t_1<t_0$ and let $(N,h)$ be a globally hyperbolic
extension of $(M,g)$.  Further, let $\mathfrak{
C}^{\prime\prime}:=\mathfrak{C}\cap\Omega^{-1}([0,t_1])$ and let us
denote by ${\cal S}^+$ the portion of $N_p$ to the future of $\Sigma
(t_1)$. Hence it is clear that $T\equiv 0$ on
$\mathfrak{C}^{\prime\prime}$. Further, let $x$ be in the
topological interior of $D^+({\cal S}^{\prime},N)$, hence
$W=J^-(x,N)\cap J^+({\cal S}^{\prime},N)$ is compact. Then $T\equiv
0$ on $W$ by the conservation theorem of Hawking and Ellis (cfr.
page 93 in $\cite{HE}$), thus $T\equiv 0$ on $\text{int}D^+({\cal
S}^{\prime},N)$. Hence by continuity we have $T\equiv 0$ on
$D^+({\cal S}^{\prime},N)\cap\tilde{M}$.

On the other hand, let $x\in
J^+(p,N)\cap\tilde{M}-\mathfrak{C}^{\prime\prime}$ and let $\gamma$
be a past inextendible timelike curve with future endpoint $x$.
Since $J^+(p,N)\cap\tilde{M}\subset D^+(N_p,N)\tilde{M}$ by lemma
$\ref{structurelemma0}$, we have that $\gamma$ must intersect $N_p$,
say at $y$. If $\Omega (y)\ge t_1$ then $y\in {\cal S}^{\prime}$. If
$\Omega (y)<t_1$ then notice that $\Omega (x)>t_1$ since
$x\not\in\mathfrak{C}^{\prime\prime}$. Now, since the function
$t\mapsto \Omega (\gamma (t))$ is continuous there exist a point
$z\in\gamma$ between $x$ and $y$ such that $\Omega (z)=t_1$. Hence
$z\in\Sigma (t_1)\subset  {\cal S}^{\prime}$. Thus we have the
inclusions $I^+(S)\subset J^+(p,N)\cap\tilde{M}\subset
\mathfrak{C}^{\prime\prime}\cup (D^+({\cal
S}^{\prime},N)\cap\tilde{M})$ where $S=\partial I^+(\eta )$ as in
lemma $\ref{structurelemma1}$. Then we just showed $T\equiv 0$ on
$I^+(S)$.

In a time dual fashion, we can show $T$ vanishes in a neighborhood
of $q$ and consequently on the whole set $I^-(S)$. To finish the
proof, recall that since $\partial I^+(\eta )=S=\partial
I^-(\eta)$ then $\tilde{M}=S\cup I^+(S)\cup I^-(S)$, therefore $T\equiv 0$ on
$\tilde{M}$ and the result follows. $\Box$

\smallskip
We conclude with a couple of remarks.  In~\cite{gal01,gal02},
a uniqueness result for Minkowski space is obtained
that is entirely analogous to theorem~\ref{rigds}.
Although, in  the asymptotically
Minkowskian setting, the fact that $\scri$ is null adds some complications to the analysis, one should still be able
to modify the techniques used here to allow a priori for the presence
of matter in that setting, as well.  Also, note that Maxwell fields are excluded
from theorem~\ref{matterfield}; they do not satisfy condition {\bf C}.  Nonetheless,
by taking advantage of the conformal invariance of such fields,
it may be possible to obtain a version of theorem \ref{matterfield} that includes them.

\section*{Acknowledgements} We would like to thank Helmut Friedrich for discussions
during the early stages of this work.  This work was supported in part by NSF grant
DMS-0405906.

%%%%%%%%%%%%%%%%%%%%%%%%%%%%%%%%%%%%%%%%%%%%%%%%%%%%%55
%%%%%%%%%%%%%%%%%%%%%%%%%%%%%%%%%%%%%%%%%%%%%%%%%%%%%%%
%%%%%%%%%%%%%%%%%%%%%%%%%%%%%%%%%%%%%%%%%%%%%%%%%%%%%%%

\end{document}